\documentclass[fleqn,usenatbib]{mnras}

\usepackage{newtxtext,newtxmath}

\usepackage[T1]{fontenc}

\DeclareRobustCommand{\VAN}[3]{#2}
\let\VANthebibliography\thebibliography
\def\thebibliography{\DeclareRobustCommand{\VAN}[3]{##3}\VANthebibliography}

\usepackage{graphicx}	
\usepackage{amsmath}	
\usepackage{natbib}
\usepackage{environ}
\usepackage{xcolor}
\usepackage{longtable}
\usepackage{array}
\usepackage{orcidlink}
\usepackage{lineno}
\usepackage{threeparttable}

\def\aj{{AJ}}%
\def\apj{{ApJ}}%
\def\apjl{{ApJ}}%
\def\apjs{{ApJS}}%
\def\aap{{A\&A}}%
\def\pasp{{PASP}}%
\def\nat{{Nature}}%
\def\mnras{{MNRAS}}%

\newcommand{\kms}{\,km\,s$^{-1}$}
\newcommand{\mstellar}{\ensuremath{M_{\mathrm{stellar}}}}
\newcommand{\omatter}{\ensuremath{\Omega_{\mathrm{M}}}}

\newcommand{\caiihk}{\ensuremath{\mathrm{Ca}\,\textsc{ii}\,\mathrm{H\&K}}}
\newcommand{\caiinir}{\ensuremath{{\mathrm{Ca}\,\textsc{ii\,NIR}}}}
\newcommand{\feiii}{\ensuremath{\mathrm{Fe}\,\textsc{iii}\,\lambda5129}}
\newcommand{\feiiineb}{\ensuremath{\mathrm{Fe}\,\textsc{ii}\,\lambda7155}}
\newcommand{\mgii}{\ensuremath{\mathrm{Mg}\,\textsc{ii}}}

\newcommand{\Siiif}{\ensuremath{\mathrm{Si}\,\textsc{ii}\,\lambda4000}}
\newcommand{\Siiihk}{\ensuremath{\mathrm{Si}\,\textsc{ii}\,\lambda3858}}
\newcommand{\Siii}{\ensuremath{\mathrm{Si}\,\textsc{ii}\,\lambda6355}}
\newcommand{\vsiii}{\ensuremath{v_{\mathrm{Si}\,\textsc{ii}}}}

\newcommand{\deltam}{\ensuremath{\Delta m_{15}}}

\title[PS1-MDS SN~Ia velocity]{Measuring the Ejecta Velocities of Type Ia Supernovae from the Pan-STARRS1 Medium Deep Survey}

\author[Y.-C.~Pan et al.]{
Y.-C.~Pan,$^{1}$\thanks{E-mail: ycpan@astro.ncu.edu.tw},
Y.-S.~Jheng$^{1}$,
D.~O.~Jones$^{2}$,
I.-Y.~Lee$^{1}$,
R.~J.~Foley$^{3}$,
R.~Chornock$^{4}$,
D.~M.~Scolnic$^{5}$,
\newauthor
E.~Berger$^{6}$,
P.~M.~Challis$^{6}$,
M.~Drout$^{7}$,
M.~E.~Huber$^{8}$,
R.~P.~Kirshner$^{6}$,
R.~Kotak$^{9}$,
R.~Lunnan$^{10}$,
\newauthor
G.~Narayan$^{11}$,
A.~Rest$^{12,13}$,
S.~Rodney$^{14}$ and
S.~Smartt$^{15}$
\\
$^{1}$Graduate Institute of Astronomy, National Central University, 300 Jhongda Road, 32001 Jhongli, Taiwan\\
$^{2}$Gemini Observatory, NSF's NOIRLab, 670 N. A'ohoku Place, Hilo, Hawai'i, 96720, USA\\
$^{3}$Department of Astronomy and Astrophysics, University of California, Santa Cruz, CA 92064, USA\\
$^{4}$Department of Astronomy, University of California, Berkeley, CA 94720-3411, USA\\
$^{5}$Department of Physics, Duke University, Durham, NC 27708, USA\\
$^{6}$Center for Astrophysics $\vert$ Harvard \& Smithsonian, Cambridge, MA 02138, USA\\
$^{7}$Department of Astronomy and Astrophysics, University of Toronto, 50 St. George St., Toronto, Ontario, M5S 3H4, Canada\\
$^{8}$Institute for Astronomy, University of Hawaii, 2680 Woodlawn Drive, Honolulu, HI 96822, USA\\
$^{9}$Department of Physics and Astronomy, University of Turku, FI-20014 Turku, Finland\\
$^{10}$Department of Astronomy, The Oskar Klein Centre, Stockholm University, AlbaNova, SE-106 91 Stockholm, Sweden\\
$^{11}$Department of Astronomy, University of Illinois at Urbana-Champaign, Urbana, IL 61801, USA\\
$^{12}$Department of Physics and Astronomy, Johns Hopkins University, 3400 North Charles Street, Baltimore, MD 21218, USA\\
$^{13}$Space Telescope Science Institute, 3700 San Martin Drive, Baltimore, MD 21218, USA\\
$^{14}$Department of Physics and Astronomy, University of South Carolina, 712 Main St., Columbia, SC 29208, USA\\
$^{15}$Astrophysics Research Centre, School of Mathematics and Physics, Queen’s University, Belfast BT7 1NN, UK\\
}

\date{Accepted XXX. Received YYY; in original form ZZZ}

\pubyear{2024}

\begin{document}
\label{firstpage}
\pagerange{\pageref{firstpage}--\pageref{lastpage}}
\maketitle

\begin{abstract}
There is growing evidence that Type Ia supernovae (SNe~Ia) may originate from multiple explosion channels. Previous studies have indicated that the ejecta velocity of SNe~Ia is one powerful tool to discriminate between different channels. In this work, we study $\sim400$ confirmed SNe~Ia discovered by the Pan-STARRS1 Medium Deep Survey (PS1-MDS), and obtain a sample of $\sim50$ SNe~Ia that have near-peak \Siii\ velocity (\vsiii) measurements. We investigate the relationships between \vsiii\ and various parameters, including SN light-curve width, color, host-galaxy properties, and redshift. No significant trends are identified between \vsiii\ and light-curve parameters. Regarding the host-galaxy properties, we see a significant trend that high-velocity (HV) SNe~Ia ($\vsiii \ga 12000$\,\kms) tend to reside in more massive galaxies compared to normal-velocity (NV) SNe~Ia ($\vsiii < 12000$\,\kms) when combining both the PS1-MDS dataset and those from previous low-$z$ studies. While we do not see a significant trend between \vsiii\ and redshift, HV SNe~Ia appear to be more prevalent in low-$z$ samples than in high-$z$ samples. We discuss several possibilities that could potentially contribute to this trend. Furthermore, we investigate the potential bias on SN~Ia distances and find no significant difference in Hubble residuals between HV and NV subgroups.
\end{abstract}




\section{Introduction}
Type Ia supernovae (SNe Ia) are mature standardizable candles that are frequently used to measure the cosmological parameters \citep[e.g.,][]{1998AJ....116.1009R,1999ApJ...517..565P,2007ApJ...659...98R,2009ApJS..185...32K,2011ApJ...737..102S,2012ApJ...746...85S,2018ApJ...859..101S,2019ApJ...872L..30A,2019ApJ...881...19J,2022ApJ...938..110B}. These studies generally assumed that the SNe~Ia originated from a uniform class and showed negligible evolution over the cosmological time. In practice, we do not fully understand these exceptional explosions, such as their progenitor system and the explosion mechanism. Thus, the degree to which SN Ia properties change with redshift and how that will impact the precision of the cosmological parameters will become an important consideration in their future use. 

Previous studies have shown some evidence that SN~Ia properties could change with redshift. \citet{2007ApJ...667L..37H} first studied the light-curve width as a function of redshift. After carefully selecting their sample to minimize the Malmquist bias, they found SNe~Ia at higher redshift tend to have a wider light-curve width. This further implies that the higher-$z$ SNe~Ia are on average brighter than the lower-$z$ SNe~Ia \citep[based on the width and luminosity relation;][]{1993ApJ...413L.105P}. This trend has also been confirmed by \citet{2021A&A...649A..74N} with a more recent dataset.

The SN~Ia spectral features as a function of redshift were also investigated by several studies. \citet{2008A&A...477..717B} studied three spectral features (\caiihk, \Siiif, \mgii) and found that the low-$z$ and high-$z$ samples are generally similar except for the equivalent width (EW) of \mgii\ line. In contrast, \citet{2011MNRAS.410.1262W} did not see a significant difference in the EW of \mgii\ between the low-$z$ and high-$z$ samples; however, they found the strength of \Siiif\ decreases with redshift. \citet{2008ApJ...684...68F} constructed the SN~Ia composite spectra based on the redshift. Their results revealed an ultraviolet (UV) excess for the composite spectrum of the high-$z$ sample, also a significant difference in the strength of \feiii\ between low-$z$ and high-$z$ composite spectra. The trend that higher-$z$ SNe~Ia tend to have a UV excess has also been confirmed by \citet{2018A&A...614A.134B}. In addition, the results from \citet{2009ApJ...693L..76S} and \citet{2009A&A...507...85B} both indicated that the higher-$z$ SNe~Ia tend to show lower intermediate-mass element (IME) abundances than that of lower-$z$ counterparts, although \citet{2009ApJ...693L..76S} attributed this trend as changes in SN~Ia demographics rather than the evolution of photometrically similar SNe.

Most studies mentioned above could not investigate the SN spectral properties redward of 6000\,\AA\ (e.g., the \Siii\ line) due to the limited wavelength coverage at high redshift. \citet{2012ApJ...748..127F} was the first to look into the relation between \Siii\ velocity (\vsiii) and redshift, but they found an insignificant difference in \vsiii\ between low-$z$ and high-$z$ SNe~Ia with a small sample. More recent studies suggested that SN~Ia ejecta velocity is likely to correlate with the host-galaxy environment. For example, high-velocity (HV) SNe~Ia \citep[$\vsiii \ga 12000$\,\kms;][]{2009ApJ...699L.139W} tend to reside in more massive galaxies than their normal-velocity (NV; $\vsiii \la 12000$\,\kms) counterparts \citep{2015MNRAS.446..354P,2020ApJ...895L...5P,2021ApJ...923..267D}. Considering the evolution of galaxy populations, \citet{2020ApJ...895L...5P} suspected that the fraction of HV SN~Ia may decrease with redshift. With a new light-curve model, \citet{2022arXiv220905584J} found a marginal trend that SNe~Ia tend to have larger \Siii\ EWs in low-mass galaxies, but they did not see a trend with \vsiii.

\citet{2009ApJ...699L.139W} proposed that HV SNe~Ia are likely to have a different origin from NV SNe~Ia, given their preference for a lower extinction ratio ($R_{V}$) than that of NV SNe~Ia. There is also evidence that the HV SNe~Ia tend to show a blue excess in their late-time light curves and variable Na\,\textsc{i} absorption lines in their spectra \citep{2019ApJ...882..120W}. In addition to their preference for massive host galaxies \citep[e.g.,][]{2015MNRAS.446..354P,2020ApJ...895L...5P}, they also tend to be more concentrated in the inner regions of their host galaxies, whereas the NV events span a wider range of radial distances \citep{2013Sci...340..170W}. In the nebular phase, a few studies have indicated that HV SNe~Ia tend to show redshifted [\feiiineb] line, while NV SNe~Ia could have either blueshifted or redshifted [\feiiineb] line in their spectra \citep{2010Natur.466...82M,2018MNRAS.477.3567M,2021ApJ...906...99L}. This trend implies that HV SNe~Ia (and at least part of the NV SNe~Ia) could be produced via asymmetric explosions. It is not yet clear if the HV and NV SNe~Ia are physically distinct and produced via different channels. Theoretical studies proposed that the sub-Chandrasekhar-mass WD explosions could be a promising mechanism to explain the wide range of \vsiii\ (including the HV SNe~Ia) as found by spectroscopic observations \citep[e.g.,][]{2019ApJ...873...84P,2021ApJ...922...68S}. The simulation of sub-Chandrasekhar-mass explosions from these studies are generally consistent with the observational properties of HV SNe~Ia, such as their unique color and ejecta velocity. However, more precise physics, radiative transfer, and the effect of secondary star \citep[e.g.,][]{2022MNRAS.tmp.2892P}  need to be considered in future works. 

Motivated by the growing evidence that SNe~Ia are likely to result from multiple channels, and that their ejecta velocities could be key to distinguishing between different channels of explosion, we measure the ejecta velocity from the \Siii\ absorption line of the SNe~Ia discovered by the Pan-STARRS1 Medium Deep Survey (PS1-MDS) in this paper. We aim to investigate the relations between \vsiii\ and several SN and host-galaxy properties, and compare them with low-$z$ studies. A plan of the paper follows. In Section~\ref{sec:data} we discuss our SN Ia spectroscopic sample, \vsiii\ measurement, and the determination of host-galaxy parameters. The results are shown in Section~\ref{sec:results}. The discussion and conclusions are presented in Sections~\ref{sec:discussion} and \ref{sec:conclusions}, respectively. Throughout this paper, we assume $\mathrm{H_0}=70$\,km\,s$^{-1}$\,Mpc$^{-1}$ and a flat universe with $\omatter=0.3$ if not mentioned otherwise.

\section{Data and method}
\label{sec:data}

\subsection{Spectroscopic observations of PS1-MDS SNe}
\label{sec:ps1-data}
Here we study the spectroscopically confirmed SNe Ia discovered by the PS1-MDS. The details of the PS1-MDS and its selection process can be found in \citet{2014ApJ...795...44R} and \citet{2018ApJ...859..101S}. The spectroscopic observations were obtained with a variety of telescope/instrument combinations, including the FAST Spectrograph on the 1.5-m Tillinghast Telescope \citep{1998PASP..110...79F}, the Supernova Integrated Field Spectrograph \citep[SNIFS;][]{2007AAS...210.8207A} on the University of Hawai'i 88-inch (UH88) telescope, the Alhambra Faint Object Spectrograph and Camera (ALFOSC) on the 2.56-m Nordic Optical Telescope, the Dual Imaging Spectrograph (DIS) on the Astrophysical Research Consortium 3.5-meter Telescope, the Richey-Chretien Spectrograph on the Kitt Peak National Observatory 4-m telescope, the Intermediate dispersion Spectrograph and Image System (ISIS) on the 4.2-m William Herschel Telescope (WHT), the Blue Channel Spectrograph \citep{1989PASP..101..713S} and Hectospec \citep{2005PASP..117.1411F} on the 6.5-m Multiple Mirror Telescope (MMT), the Low Dispersion Survey Spectrograph-3 (LDSS3) and the Magellan Echellette \citep[MagE;][]{2008SPIE.7014E..54M} on the 6.5-m Magellan Clay telescope, the Inamori-Magellan Areal Camera and Spectrograph \citep[IMACS;][]{2011PASP..123..288D} on the 6.5-m Magellan Baade telescope, the Gemini Multi-Object Spectrographs \citep[GMOS;][]{2004PASP..116..425H} on both 8.1-m Gemini North and South telescopes, the DEep Imaging Multi-Object Spectrograph (DEIMOS) on the 10-m Keck-II telescope \citep{2003SPIE.4841.1657F}, and the Optical System for Imaging and low-Intermediate-Resolution Integrated Spectroscopy \citep[OSIRIS;][]{2003SPIE.4841.1739C} on the 10.4-m Gran Telescopio CANARIAS (GTC).

The spectra are reduced with IRAF\footnote{The Image Reduction and Analysis Facility (IRAF) is distributed by the National Optical Astronomy Observatories, which are operated by the Association of Universities for Research in Astronomy, Inc., under a cooperative agreement with the National Science Foundation.} and Interactive Data Language (IDL) tasks following the standard procedure (bias subtraction, flat-fielding, wavelength calibration, and flux calibration). When possible, we carefully select the background regions to subtract the host-galaxy light when extracting the 1-D spectrum. This can help reduce the host-galaxy contamination on the SN spectra. A summary of spectroscopic observation for all the confirmed SNe Ia can be found in Table~\ref{spec_sample1}, and their spectra are available from the WISeREP archive \citep{2012PASP..124..668Y}.

\begin{table}
\centering
\caption{Summary of spectroscopically confirmed SN~Ia in this work. The entire table is available online as supplementary material. A portion is shown here for guidance regarding its form and content.}
\label{spec_sample1}
\begin{tabular}{lccclc}
\hline\hline
SN Name & Telescope & MJD & Phase (days) & $z_{\rm CMB}$ & $z$ source\\
\hline
PS1-909006 & Gemini & 55098 & 17.1 & 0.283 & host \\
PS1-909010 & Gemini & 55105 & 6.4 & 0.276 & host \\
PS1-910016 & Gemini & 55125 & 19.7 & 0.22 & SN \\
PS1-910017 & Gemini & 55125 & 12.0 & 0.32 & SN \\
PS1-910018 & Gemini & 55125 & 12.1 & 0.264 & host \\
PS1-910020 & Gemini & 55124 & 14.3 & 0.242 & host \\
PS1-910021 & Gemini & 55125 & 20.2 & 0.256 & host \\
PS1-000006 & Magellan & 55216 & 10.0 & 0.231 & host \\
PS1-000010 & Magellan & 55217 & 1.7 & 0.246 & host \\
PS1-000011 & Gemini & 55214 & 2.7 & 0.381 & host \\
\hline
\end{tabular}
\end{table}

\begin{table}
\centering
\caption{The SN sample selection in this work.}
\begin{tabular}{lcc}
\hline\hline
 & Num. of SNe left\\
\hline
Confirmed as SN~Ia & 381 \\
Redshift derived from host & 321 \\
LC quality cut & 283 \\
Phase cut & 126 \\
With \Siii\ measurement & 48 \\
\hline
\end{tabular}
\label{selection}
\end{table}

\subsection{Sample selection}
\label{sec:selection}

The objects studied in this work were all spectroscopically confirmed as SN Ia. We use the Supernova Identification \citep[SNID;][]{2007ApJ...666.1024B} code to classify the SNe. The classification scheme is briefly summarized as follows: We first clipped or masked the nebular emission lines from the host galaxy (if any) in the spectra before SNID classification. The redshift is forced to be fixed if known (e.g., from the host galaxy). The SNID classification is mainly based on the quality parameter named `\textit{rlap}' \citep[see][]{2007ApJ...666.1024B}. The higher \textit{rlap} indicates a better correlation (as weighted by the overlap in wavelength space) between the input spectrum and the template. Following the method in \citet{2007ApJ...666..674M}, we classify a supernova to be Ia if its best-match template from SNID is an Ia and at least 50\% of the templates with \textit{rlap} larger than 5.0 are also Ia. A total of 362 objects pass the criterion and are confirmed as Ia. We classify another 19 objects as Ia (which fail to pass our SNID criteria) with eye inspection. The final parent sample contains 381 spectroscopically confirmed SNe~Ia. 

To measure the ejecta velocity, only SNe with host-galaxy redshift available are included. This gives 321 SNe from the parent sample. We also apply the same quality cuts on SN light curves (LCs) as those described in \citet{2018ApJ...859..101S} to ensure a precise phase determination of our spectrum. These include the constraints on LC fitting parameters such as reduced $\chi^2$, the LC width ($x_1$), and peak date ($pkmjd$) uncertainties. A total of 283 SNe~Ia pass the LC quality cut and are used as our final spectroscopic sample. The number of SNe left after each cut is shown in Table~\ref{selection}. 

\begin{figure}
	\centering
	\begin{tabular}{c}
		\includegraphics*[scale=0.5]{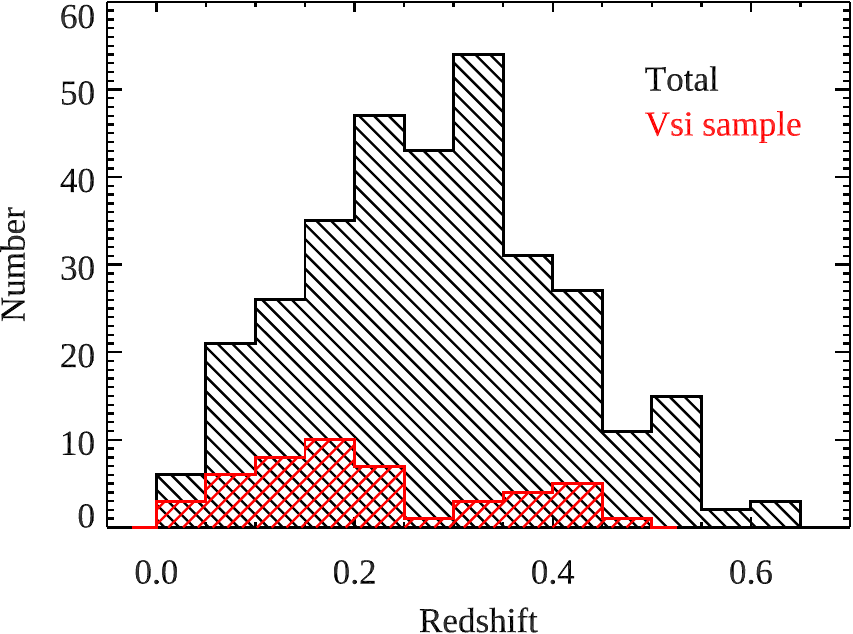}\\
		\vspace{0.25cm}
		\includegraphics*[scale=0.5]{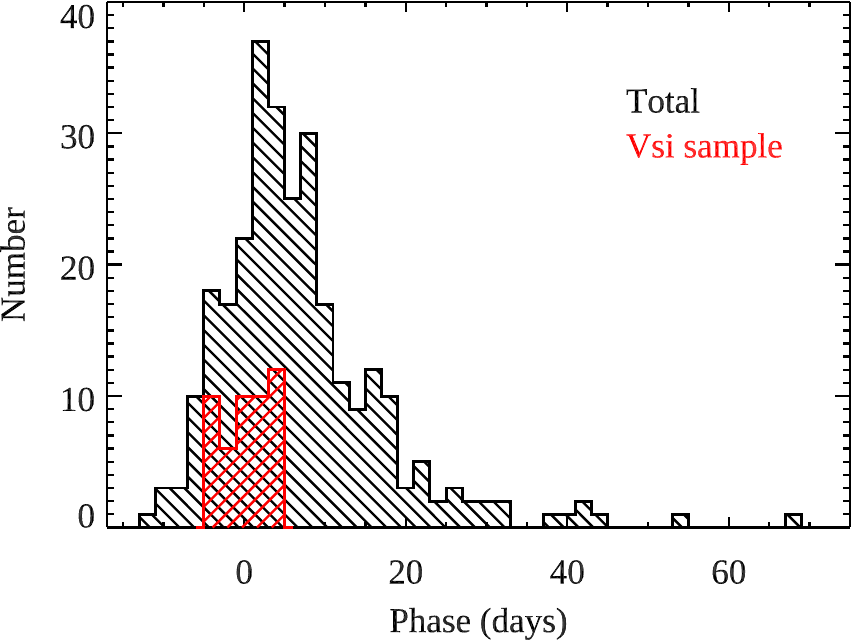}
	\end{tabular}
       \caption{\textit{Upper:} The redshift distributions of our spectroscopic sample (black histogram) and those with \vsiii\ measurements available (red histogram). \textit{Lower:} The phase distribution of the spectra we used for classification (black histogram) and those with \vsiii\ measurements available (red histogram).
      }
        \label{sample}
\end{figure}

The redshift of our final spectroscopic sample ranges from 0.03 to 0.63, with a median redshift of $\sim$0.3. Most of the SNe have only single-epoch spectroscopic observation, with $<$10\% of our SNe (21 out of 283) having multi-epoch spectra. The distribution of redshift and phase of the spectra used for classification can be found in Fig.~\ref{sample}.

\subsection{\Siii\ velocity measurement}
\label{sec:vel-measure}
In this work, we focus on the photospheric velocity measured from the \Siii\ line near the maximum light. \Siii\ line is selected because it is relatively free from the contamination of other lines and detached high-velocity features \citep[HVFs;][]{2003ApJ...591.1110W,2004ApJ...607..391G,2005ApJ...623L..37M} near the maximum light. The alternatives are the \caiihk\ and \caiinir\ lines. However, the \caiinir\ line is not covered by the optical spectrum of our high-$z$ sample. Regarding the \caiihk, the existence of HVFs and potential contamination from the \Siiihk\ line could complicate the precise determination of the photospheric component when high-quality spectra are unavailable. \citep[e.g.,][]{2013MNRAS.435..273F,2014MNRAS.437..338C,2014MNRAS.444.3258M,2015MNRAS.451.1973S}. Given the high-$z$ nature of our dataset, we only focus on measuring the \Siii\ velocity in this work.

Following \citet{2014MNRAS.444.3258M} and \citet{2015MNRAS.446..354P}, we obtain the near-peak sample by restricting the phase of our spectroscopic observations to be within 5~days relative to the peak luminosity. The velocity evolution within this phase range is expected to be mild and small \citep[e.g., with a mean gradient of $\sim30$\,\kms\ per day;][]{2015MNRAS.446..354P}. This phase cut gives a near-peak sample of 126 SNe. We further exclude 14 SNe due to the insufficient wavelength coverage on the \Siii\ feature. Following \citet{2012ApJ...748..127F}, we also perform a visual inspection to exclude the spectra that are noisy and/or have a flat absorption feature, have significant sky residuals, have obvious reduction issues, or combinations of the above factors. Finally, we are able to measure the \vsiii\ from 48 SNe~Ia in our sample.

We use the technique from \citet{2006AJ....131.1648B} to measure the velocity. Following the procedure described in \citet{2019MNRAS.486.5785S}, the spectrum is first smoothed with an inverse-variance Gaussian kernel of a temporary smoothing factor. An error spectrum is then determined from the smoothed residual spectrum by subtracting the initially smoothed spectrum from the original spectrum. An S/N is estimated using the median flux ratio between the smoothed and the error spectrum. This S/N will be used to determine the final smoothing factor \citep[by adopting the empirical relation in][]{2019MNRAS.486.5785S} to smooth the SN spectrum. The \vsiii\ is calculated by measuring the wavelength at the minimum flux of the smoothed \Siii\ line. The uncertainties in \vsiii\ are estimated by randomly varying the smoothing factor 10000 times within the typical range of our sample (0.002 to 0.0045). It is noted that the uncertainties in \vsiii\ are not considered when classifying our SNe into the HV and NV subgroups. However, we find that this has a negligible effect on our results.

The majority of SNe in our final sample has only a single near-peak spectrum, with two SNe (PS1-350083 and PS1-510266) having multiple near-peak spectra. For the two near-peak spectra of PS1-350083, we adopt the one nearest to the SN maximum light. For PS1-510266, there are also two near-peak spectra, but only one has \Siii\ available. The median \vsiii\ of our sample is 10923\,\kms, with a minimum and maximum velocity of 9015 and 13459\,\kms, respectively. The redshift distribution of our \vsiii\ sample can be found in Fig.~\ref{sample}. Fig.~\ref{vsi_mosaic1} and Fig.~\ref{vsi_mosaic2} show the observed and smoothed flux of each SN around \Siii\ line in our \vsiii\ sample. All the \vsiii\ measurements can be found in Table~\ref{hiz_vsi}.

\subsubsection{Evaluation of host-galaxy contamination}
\label{sec:host-contamination}
\begin{figure}
	\centering
	\begin{tabular}{c}
		\includegraphics*[scale=0.5]{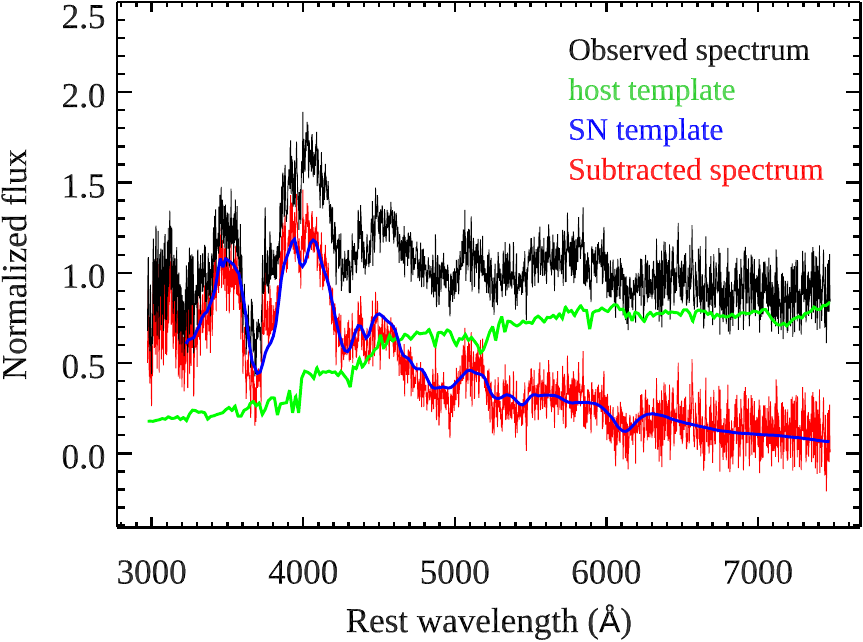}
	\end{tabular}
       \caption{Result of host-galaxy subtraction for SN~Ia PS1-190260. The back and red spectra represent the original SN spectrum and that after the host-galaxy subtraction, respectively. The green spectrum is the host-galaxy template determined from the host-galaxy photometry. The SN template produced by \texttt{kaepora} is shown in blue. 
      }
        \label{host-sub-example}
\end{figure}

\begin{figure}
	\centering
	\begin{tabular}{c}
		\includegraphics*[scale=0.5]{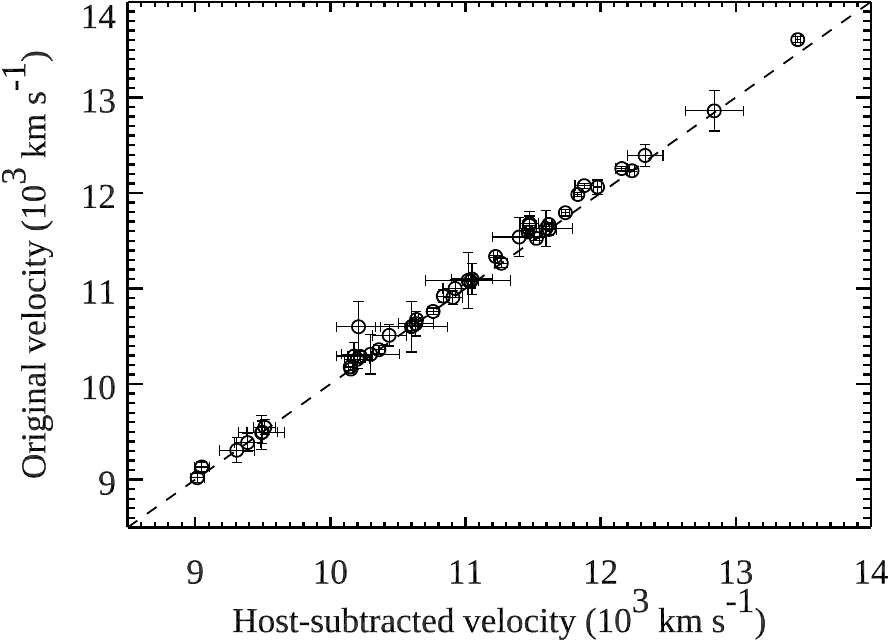}
	\end{tabular}
       \caption{The \Siii\ velocity measured from the SN spectra before and after the host-galaxy template subtraction. The line of equality is shown in a dashed line.
      }
        \label{host-sub-compare}
\end{figure}

High-$z$ SNe are often blended with their host galaxies. As mentioned in Section~\ref{sec:ps1-data}, we chose the background regions as carefully as possible to subtract the host-galaxy light when processing the spectrum. However, it is sometimes inevitable that the final spectrum could still suffer non-negligible contamination from their host galaxies. Here we evaluate the effect of host-galaxy contamination on the \vsiii\ measurement.

Our method is based on \citet{2005ApJ...634.1190H}, using a $\chi^2$ fitting technique to estimate the fraction of host-galaxy contamination in the SN spectrum. \citet{2005ApJ...634.1190H} fit each SN spectrum with a library of SN templates of all types and a range of phases. The reddening, redshift (if not determined from the host galaxy), and fraction of host-galaxy contamination are derived by minimizing the difference between the observed SN spectrum and template spectrum. Instead of fitting the observed spectrum with templates of a wide range of parameter space (which could result in substantial degeneracy), we allow only specific SN and host-galaxy templates in the fitting. For the SN template, we use the \texttt{kaepora} database \citep{2019MNRAS.486.5785S} to produce a mean spectrum given the exact phase and LC width \citep[e.g., from][]{2018ApJ...859..101S} of each SN in our sample. For the host-galaxy template, we first measure the host-galaxy photometry from the PS1 template images in $grizy$ filters. We measure the photometry by placing an aperture of 1.5\arcsec\ radius at the SN location. This allows us to better constrain the host-galaxy spectral energy distribution (SED) near the SN birthplace. The measured photometry will then be used as input for the photometric redshift code \texttt{Z-PEG} to derive the best-fit galaxy template and properties. \texttt{Z-PEG} fits the observed galaxy colors with galaxy SED templates corresponding to nine spectral types (SB, Im, Sd, Sc, Sbc, Sb, Sa, S0, and E). Here we assume a \citet{1955ApJ...121..161S} initial-mass function (IMF). The photometry is corrected for the foreground Milky Way reddening with $R_{V} = 3.1$ and a \citet*[][CCM]{1989ApJ...345..245C} reddening law. The details of the fitting procedure can be found in \citet{2014MNRAS.438.1391P}.

Fig.~\ref{host-sub-example} shows the result of host-galaxy subtraction from one of our SNe. Overall, the host-galaxy contamination of our sample is not significant. We determine a median fraction of host-galaxy contamination of only 17\%, with 45 out of 48 SNe having fractions lower than 50\%. The \vsiii\ comparison between the spectra before and after the host-galaxy template subtraction can be found in Fig.~\ref{host-sub-compare}. The two measurements are consistent with each other, with a mean offset of 65\,\kms\ and rms scatter of 101\,\kms. Given that the difference in velocity is statistically insignificant, and not all the SNe in our sample have host-subtracted spectra available, we simply adopt the original velocity measurements for further analysis. 

\begin{figure}
	\centering
	\begin{tabular}{c}
		\includegraphics*[scale=0.5]{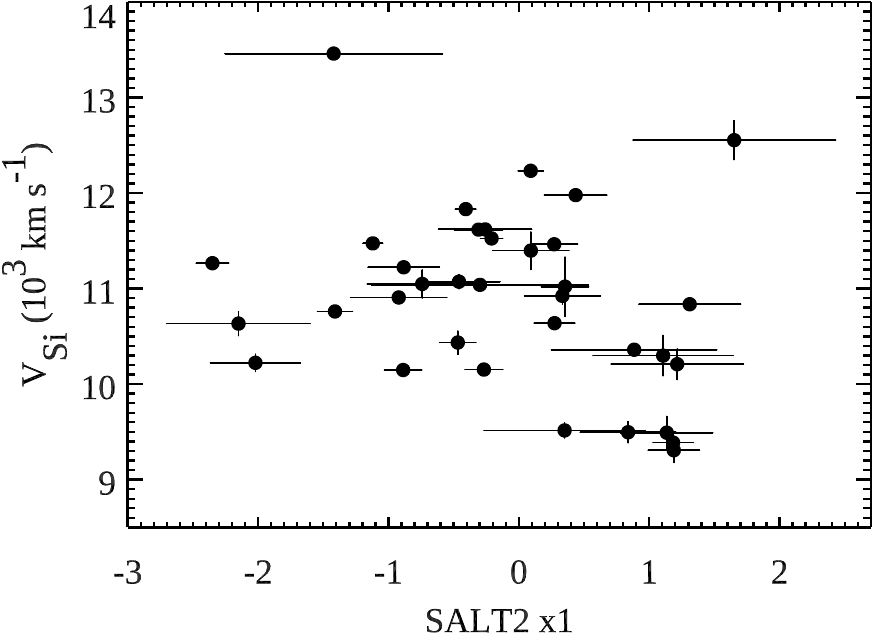}\\
		\includegraphics*[scale=0.5]{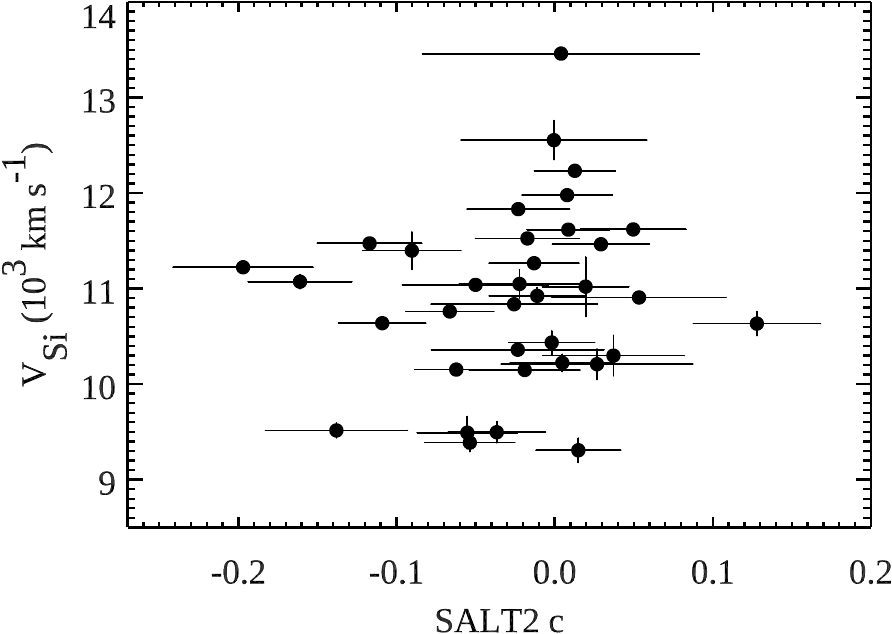}
	\end{tabular}
       \caption{The \Siii\ velocity (\vsiii) as a function of SN stretch $x_1$ (upper panel) and color $c$ (lower panel) derived by the SALT2 LC fitter \citep{2007A&A...466...11G}.
      }
        \label{vsi-lc}
\end{figure}

\begin{figure*}
	\centering
	\begin{tabular}{c}
		\includegraphics*[scale=0.65]{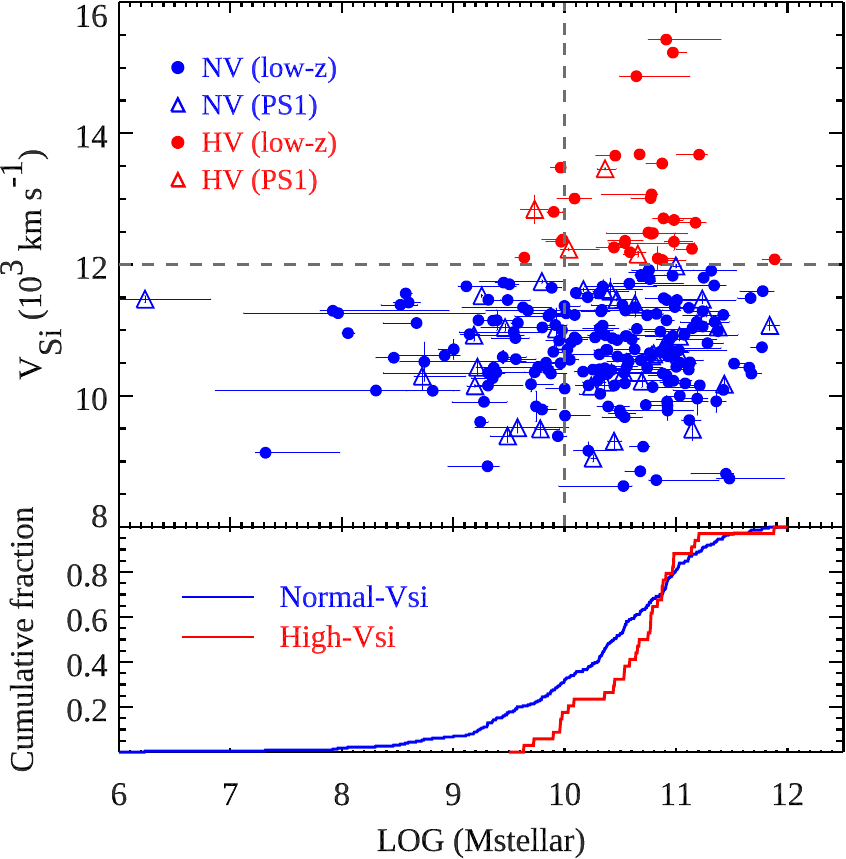}
	\end{tabular}
       \caption{The \Siii\ velocities (\vsiii) as a function of host-galaxy stellar mass (\mstellar). The PS1-MDS SNe (this work) are represented by open triangles, and the low-$z$ SNe from \citet{2020ApJ...895L...5P} are in solid circles. The HV and NV SNe~Ia in each sample are shown in red and blue, respectively. The horizontal dashed line represents the criteria for splitting the sample in velocity space. The vertical dashed line represents the host-galaxy stellar mass of $\log(\mstellar)=10\,M_{\odot}$. The bottom histograms show the cumulative fractions of \mstellar\ for HV (in red) and NV (in blue) SNe Ia.
      }
        \label{vsi-mass}
\end{figure*}

\section{Results}
\label{sec:results}
In this section, we investigate the relationship between \vsiii\ and several critical parameters. Various methods are used when evaluating the trend of our dataset, such as linear fitting with \textsc{LINMIX} \citep{2007ApJ...665.1489K} and weighted mean while comparing the difference between subgroups, if not mentioned otherwise.

\subsection{Silicon velocity and light-curve parameters}
\label{sec:vsi-lc}
We first compare the \vsiii\ with the SN photometric parameters. The SALT2 \citep{2007A&A...466...11G} stretch $x_1$ and color $c$ parameters measured in \citet{2018ApJ...859..101S} are used in the analysis. The $x_1$ parameter measures the LC width of SN~Ia, a key parameter to calibrate the SN luminosities. Brighter SNe~Ia tend to have LCs with slower decline rates or higher stretches \citep{1993ApJ...413L.105P}.  The color parameter $c$ measures the SN ($B-V$) color at maximum light, serving as another parameter in SN~Ia standardization \citep*[e.g.,][]{1996ApJ...473...88R}. Our \vsiii\ sample includes 35 SNe with $x_1$ and $c$ measurements, of which only 3 are HV SNe~Ia.

The \vsiii\ as a function of $x_1$ can be found in the upper panel of Fig.~\ref{vsi-lc}. We do not see a significant trend between \vsiii\ and $x_1$. No trend is found when comparing the HV and NV subgroups either. The results from PS1-MDS sample are generally consistent with low-$z$ studies \citep[e.g.,][]{2009ApJ...699L.139W,2014MNRAS.444.3258M,2021ApJ...923..267D}, although we note that the HV subgroup contains only 3 SNe in this analysis.

The \vsiii\ as a function of $c$ can be found in the lower panel of Fig.~\ref{vsi-lc}. We do not detect any significant trend between \vsiii\ and the color parameter $c$. Some previous studies showed that the HV SNe~Ia tend to be redder than their NV counterparts \citep[e.g.,][]{2009ApJ...699L.139W,2011ApJ...742...89F}. We note that the HV subgroup in our sample has a mean $c$ 0.03\,mag\ redder than the NV subgroup (at $3.5\sigma$ significance). However, we cannot conclusively claim the trend due to the small size of the HV subgroup (which contains only 3 HV SNe~Ia in the analysis). 

\subsection{Silicon velocity and host-galaxy properties}
\label{sec:vsi-host}
The relation between \vsiii\ and host-galaxy properties at low redshift was investigated by several recent studies \citep[e.g.,][]{2013Sci...340..170W,2015MNRAS.446..354P,2020ApJ...895L...5P}. For instance, \citet{2020ApJ...895L...5P} found a significant trend between \vsiii\ and host-galaxy stellar mass (\mstellar) with SNe~Ia discovered at $z<0.2$, in the sense that HV SNe~Ia tend to explode in massive galaxies, while NV SNe~Ia can be found in either low-mass or massive galaxies. This trend was further confirmed by \citet{2021ApJ...923..267D} with a different low-$z$ dataset. 

We revisit this relationship with the PS1-MDS sample. The host-galaxy \mstellar\ studied in this work is determined by \texttt{Z-PEG} (see Section~\ref{sec:host-contamination} for details). In contrast to Section~\ref{sec:host-contamination}, an elliptical aperture is used here to enclose the whole galaxy for measuring the global properties. Both \vsiii\ and host-galaxy stellar mass are measured consistently with those in \citet{2020ApJ...895L...5P}. Our \vsiii\ sample includes 41 SNe with host-galaxy measurements, of which only 4 are HV SNe~Ia. The result is shown in Fig.~\ref{vsi-mass}. While the trend between \vsiii\ and host-galaxy \mstellar\ is not statistically significant for the PS1-MDS sample alone, it generally shows good consistency with the low-$z$ sample. Despite the small sample size, all the HV SNe~Ia from PS1-MDS have host-galaxy $\log(\mstellar)>9.5\,M_{\odot}$, while the NV SNe~Ia can be found in galaxies with very low stellar mass (e.g., $\mstellar\sim10^{6}\,M_{\odot}$). The difference in mean $\log(\mstellar)$ between HV and NV subgroups (considering both the low-$z$ and PS1-MDS SNe) is $0.31\pm0.08\,M_{\odot}$ (at 3.9$\sigma$ significance), with HV subgroup having (on average) more massive host galaxies than that from NV subgroup.

We further examine the relationship between \vsiii\ and other host-galaxy properties, including the star-formation rate (SFR) and specific SFR. Our analysis of the PS1-MDS sample does not reveal any significant trends, which is consistent with findings from previous low-$z$ studies \citep[e.g.,][]{2020ApJ...895L...5P}. No significant trend is identified either when combining both the PS1-MDS and low-$z$ datasets. 

\subsection{Silicon velocity and redshift}
\label{sec:vsi-evolution}

\begin{figure*}
	\centering
	\begin{tabular}{c}
		\includegraphics*[scale=0.59]{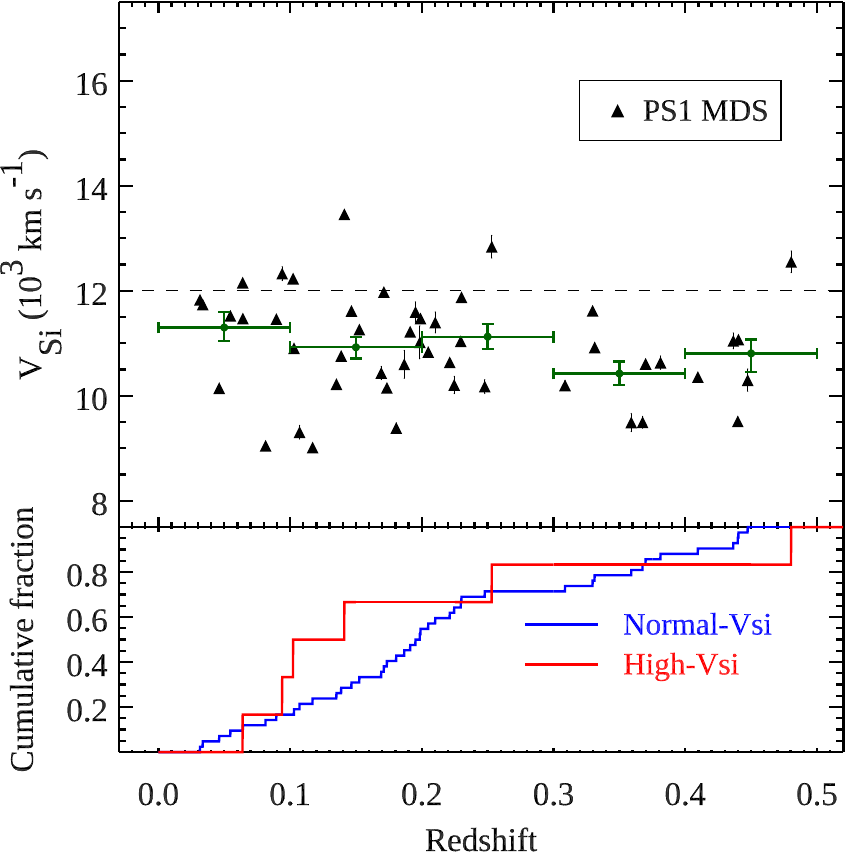}
            \hspace{0.25cm}
            \includegraphics*[scale=0.59]{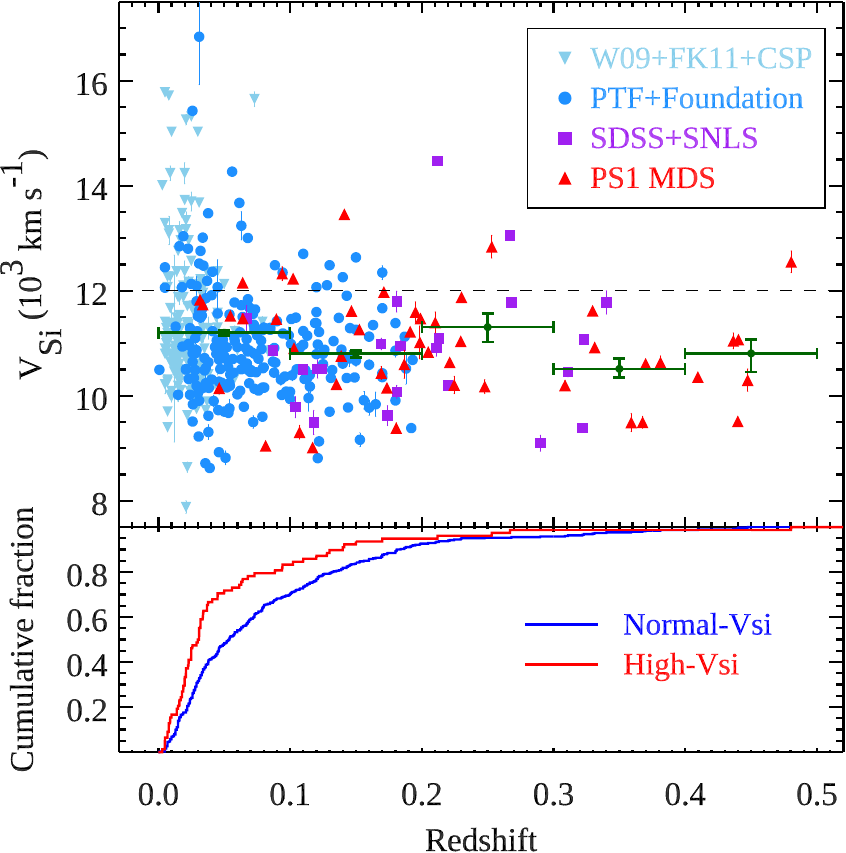}
	\end{tabular}
       \caption{\textit{Left:} The \Siii\ velocities (\vsiii) as a function of redshift for PS1-MDS sample. The PS1-MDS sample (this work) is shown in black triangles.  The mean \vsiii\ in bins of redshift is shown in green. The bottom histograms show the cumulative fractions of redshift for HV (in red) and NV (in blue) SNe Ia. \textit{Right:} The same as the left panel, but combining the SNe~Ia found in PS1-MDS (in red triangles), SDSS/SNLS (in purple squares), PTF/Foundation (in blue circles), and W09/FK11/CSP (in blue triangles).
      }
        \label{vsi-redshift}
\end{figure*}

We next investigate the relationship between \vsiii\ and redshift. \citet{2012ApJ...748..127F} compared the \vsiii\ distributions between low-$z$ and high-$z$ samples but found no significant difference between the two groups. We further explore this with our PS1-MDS sample. As shown in the left panel of Fig.~\ref{vsi-redshift}, we do not see a significant trend between \vsiii\ and redshift when considering only the PS1-MDS sample. Fitting a straight line using the \textsc{LINMIX} gives a 73\% probability the slope is negative. In addition, no discrepancy is observed in terms of the mean redshift when separating the PS1-MDS sample into NV and HV subgroups. The difference in mean redshift calculated via bootstrap resampling between HV and NV subgroups is $0.03\pm0.05$ (at $<1\sigma$ significance). However, we note that the mean value of the HV subgroup could be uncertain due to the small sample size of the PS1-only sample.

To further constrain the trend, we add the spectroscopic compilation studied in \citet{2012ApJ...748..127F} to increase the number of high-$z$ SNe. This compilation comprises the SNe discovered by both Sloan Digital Sky Survey-II (SDSS-II) Supernovae Survey \citep{2008AJ....135..338F} and Supernova Legacy Survey \citep[SNLS;][]{2006A&A...447...31A} programs. We re-measured the \vsiii\ for these SNe with the consistent method as that described in Section~\ref{sec:vel-measure}. The same phase cut (see Section~\ref{sec:selection}) is adopted to these SNe to obtain the near-peak measurements. A total of 23 high-$z$ SNe are added to our sample and listed in Table~\ref{hiz_vsi}. To include the low-$z$ sample, we use the spectroscopic compilation studied in \citet{2015MNRAS.446..354P} and \citet{2021ApJ...923..267D}. The SNe studied in \citet{2015MNRAS.446..354P} were all discovered by the Palomar Transient Factory \citep[PTF;][]{2009PASP..121.1395L}. The \citet{2021ApJ...923..267D} sample comprises low-$z$ SNe~Ia from the Carnegie Supernova Project \citep[CSP;][]{2006PASP..118....2H}, \citet[][hereafter W09]{2009ApJ...699L.139W}, \citet[][hereafter FK11]{2011ApJ...729...55F}, and Foundation Supernova Survey \citep{2018MNRAS.475..193F}. The \vsiii\ of these low-$z$ SNe were measured consistently. These low-$z$ samples contain 388 SNe~Ia with near-peak \vsiii\ measurements. 

The result is shown in the right panel of Fig.~\ref{vsi-redshift}. With such a large compilation, we still find no significant trend between \vsiii\ and redshift up to $z\sim0.5$. \textsc{LINMIX} gives only a 68\% probability the slope is negative when fitting a straight line to the whole sample. However, it is worth noting that the mean \vsiii\ of SNe in the $z<0.1$ bin is significantly higher than that for the $0.1<z<0.5$ bin (at $4.4\sigma$ significance). When separating our sample into HV and NV subgroups, we also find some evidence that the HV SNe~Ia are more likely to explode in the low-$z$ Universe than their NV counterparts. The difference in mean redshift between HV and NV subgroups is $0.027\pm0.007$ (at 3.9$\sigma$ significance), with the HV subgroup having (on average) lower redshift than the NV subgroup. A K-S test gives a $p$-value of $8\times10^{-5}$, rejecting the null hypothesis that the NV and HV SNe~Ia have the same redshift distributions.

Since no obvious trend is identified between \vsiii\ and redshift for the PS1-only sample (or even a larger sample combining all the high-$z$ SNe from PS1, SDSS, and SNLS), we suspect that the trend between HV and NV SNe~Ia found above could be due to the inclusion of low-$z$ samples. While the smaller sample size of our high-$z$ sample may play a role, the different selections and strategies between low-$z$ and high-$z$ surveys could complicate our interpretation and thus need careful consideration. A detailed discussion on this can be found in Section~\ref{sec:cause}.

\section{Discussion}
\label{sec:discussion}

\subsection{Silicon velocity distribution}
\label{sec:populations}

\citet{2020MNRAS.499.5325Z} studied the \vsiii\ distribution with a large sample of low-$z$ SNe~Ia. They found that the \vsiii\ shows a significant bimodal Gaussian distribution, with one group having a mean $\vsiii=11000\pm700$\,\kms\ and the other group having a mean $\vsiii=12300\pm1800$\,\kms. Here we designate the SNe in these two velocity distributions as group-I (with lower mean \vsiii) and group-II (with higher mean \vsiii) to distinguish them from the conventional NV and HV SNe~Ia. \citet{2020MNRAS.499.5325Z} calculated a population fraction of 65\% and 35\% for group-I and group-II SNe, respectively. While it is unclear if these populations are physically distinct, a bimodal Gaussian distribution can adequately describe the \vsiii\ distribution of low-$z$ SNe~Ia.

However, the results from \citet{2020MNRAS.499.5325Z} could be strongly biased, as a substantial fraction of their sample were discovered from the galaxy-targeted surveys \citep[from the \texttt{kaepora} database;][]{2019MNRAS.486.5785S}. Their result may not represent the intrinsic population ratio of group-II to group-I SNe, given that the HV SNe~Ia (predominantly in group-II) are preferentially discovered in massive galaxies. Here we revisit their works with low-$z$ un-targeted samples, including SNe from PTF and Foundation surveys. To facilitate a better comparison with \citet{2020MNRAS.499.5325Z}, we apply a redshift cut on these samples and use only SNe~Ia at $z<0.09$. 

\begin{figure*}
	\centering
	\begin{tabular}{c}
		\includegraphics*[scale=0.55]{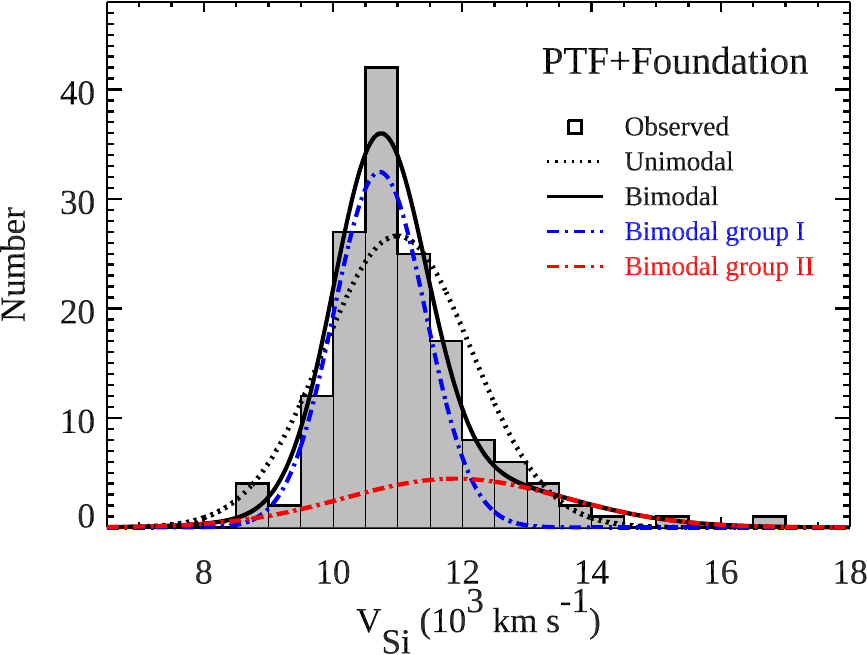}
		\hspace{0.3cm}
		\includegraphics*[scale=0.55]{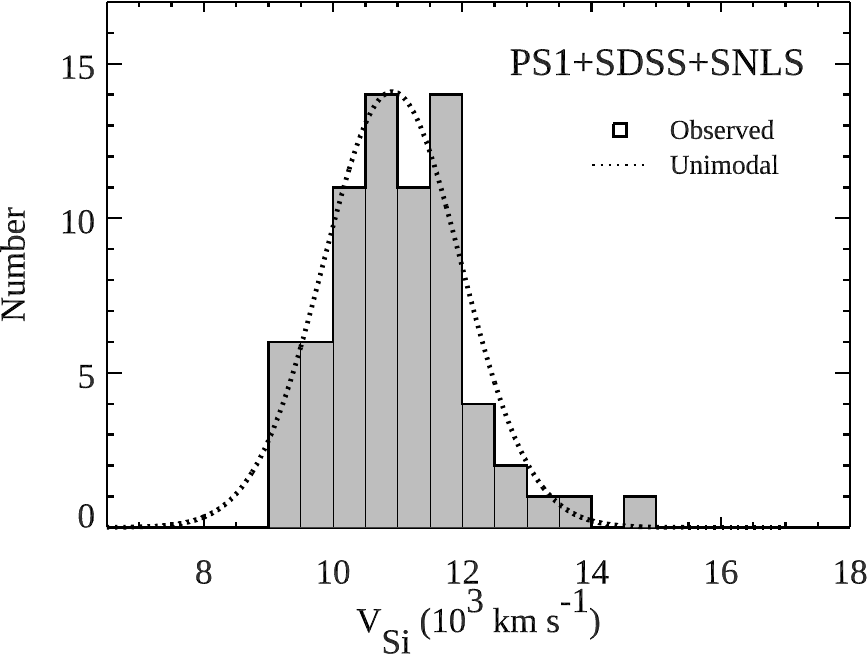}
	\end{tabular}
       \caption{\textit{Left}: The histogram shows the \Siii\ velocity (\vsiii) distribution from the low-$z$ PTF \citep{2020ApJ...895L...5P} and Foundation \citep{2021ApJ...923..267D} SNe~Ia. The black dotted and solid curves represent the unimodal and bimodal fits from the maximum-likelihood method \citep{2020MNRAS.499.5325Z}, respectively. The blue and red dash-dotted curves represent the two components in the bimodal fit. \textit{Right}: The same as the left panel, but with the high-$z$ PS1-MDS SNe (this work) and SDSS/SNLS SNe from \citet{2012ApJ...748..127F}. Here only the unimodal distribution can provide a reasonable fit to the high-$z$ dataset.
      }
        \label{population}
\end{figure*}

We repeat the same statistical technique (the maximum-likelihood method) as that performed in \citet{2020MNRAS.499.5325Z}. This technique has the advantage that it is not sensitive to the bin size of the distribution during the fitting. The result is shown in the left panel of Fig.~\ref{population}. We find a similar bimodal \vsiii\ distribution with our sample. A modified version of Akaike Information Criterion \citep[AICc;][]{1978Sugiura} of 2580 and 2551 are calculated for unimodal fit and bimodal fit of our \vsiii\ distribution, respectively. This indicates that the bimodal distribution gives a significantly better \citep[e.g., a difference larger than 6;][]{1998ApJ...508..314M} fit than the unimodal fit. However, our low-$z$ SNe tend to show quite different population fractions and mean velocities between the two groups from that of \citet{2020MNRAS.499.5325Z}. We measure a mean $\vsiii=10700\pm700$\,\kms\ and $\vsiii=11900\pm1700$\,\kms\ and population fractions of 76\% and 24\%, for group-I and group-II SNe, respectively. Our sample has an averagely lower \vsiii\ (for both velocity groups), and a lower fraction of SNe in group-II than that of \citet{2020MNRAS.499.5325Z}. We believe this should be closer to the intrinsic \vsiii\ distribution.

We next examine the \vsiii\ distribution for high-$z$ SNe~Ia. The high-$z$ sample studied here is the same as that studied in Section~\ref{sec:vsi-evolution} (from the PS1-MDS, SDSS, and SNLS). The result is shown in the right panel of Fig.~\ref{population}. In contrast to the low-$z$ sample, we find the bimodal distribution fails to provide a reasonable fit to our high-$z$ sample (we therefore only present the unimodal fit in the figure). An AICc of 1198 and 1199 is calculated for unimodal fit and bimodal fit, respectively. This indicates that the bimodal distribution is not favored over the unimodal distribution. A mean $\vsiii=10900\pm1100$\,\kms\ is determined from the unimodal fit.  

It is unclear if the difference in \vsiii\ distributions between the low-$z$ and high-$z$ samples is due to the relatively small size of the current high-$z$ sample. It is also worth noting that the conventional criterion to separate the HV and NV SNe~Ia (i.e., $\vsiii\sim12000$\,\kms) cannot separate the two velocity groups unambiguously. There is a substantial overlap in the velocity space between group-I and group-II SNe. From the bimodal fit of our low-$z$ sample, there is 3.5\% of SNe in group-I that are classified as HV SN~Ia and 52.5\% of SNe in group-II that are classified as NV SNe~Ia. Studies on each individual group of SNe using this velocity criterion would unavoidably suffer some contamination from the other group. Including additional parameters (such as carbon and nebular spectral features) could be helpful in breaking the degeneracy in velocity space.

 The large dispersion of \vsiii\ found in group-II SNe could be linked to the detonations of sub-Chandrasekhar-mass WDs based on a few theoretical studies. \citet{2019ApJ...873...84P} studied the 1D sub-Chandrasekhar-mass WD explosion models and found they can produce a wide range of \vsiii, including both HV and NV SNe~Ia. Their results showed that the relation between SN~Ia luminosity and \vsiii\ could be well explained by two populations, with one associated with the Chandrasekhar-mass WD explosions and the other with the sub-Chandrasekhar-mass WD explosions. In contrast, \citet{2021ApJ...922...68S} used multi-dimensional radiation transport calculations and produced the bulk of observed SNe~Ia from simply the double detonations of sub-Chandrasekhar-mass WDs. However, it is unclear if this scenario alone can explain the bimodal \vsiii\ distribution found for low-$z$ SNe~Ia. A large sample with well-measured ejecta velocities would be critical to better constrain the models.

\subsection{The redshift distributions of HV and NV SNe~Ia}
\label{sec:cause}
\begin{figure}
	\centering
	\begin{tabular}{c}
		\includegraphics*[scale=0.55]{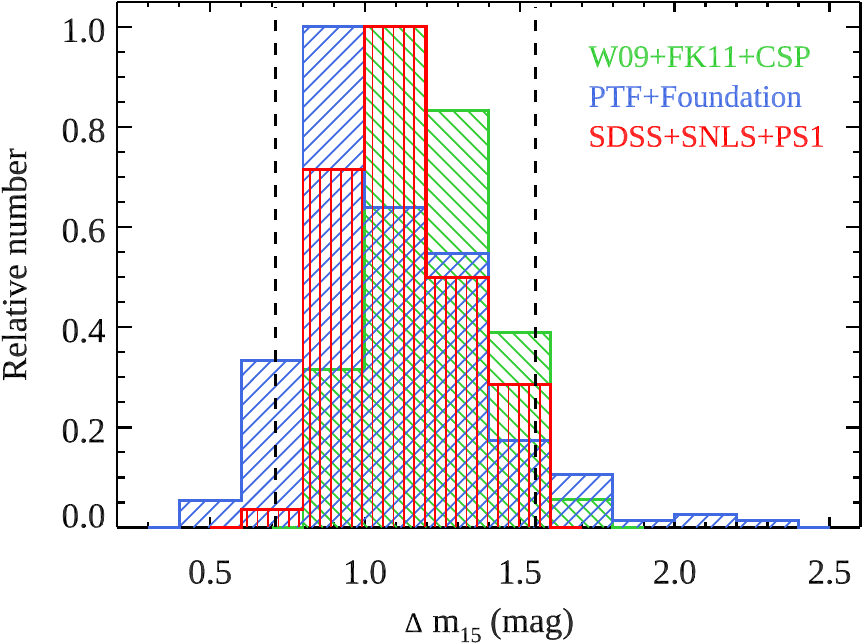}
	\end{tabular}
       \caption{The \deltam\ distributions of different sub-samples (grouped in terms of their redshifts). The vertical dashed lines represent the minimum and maximum \deltam\ that the HV SNe~Ia have in this work.
      }
        \label{dm15-hist}
\end{figure}

In Section~\ref{sec:vsi-evolution}, we investigate the relation between \vsiii\ and redshift but find no significant trend with either the PS1-only sample or a large spectroscopic compilation. However, we find the HV SNe~Ia tend to show a notable difference in redshift distribution compared to NV SNe~Ia, in the sense that the HV SNe~Ia are more prevalent at low-$z$. Since this trend vanishes when considering solely the high-$z$ samples, we suspect that the discrepancy between the low-$z$ and high-$z$ surveys (such as selection effects) could be the underlying factor driving the trend.

Firstly, the Malmquist bias may have some impact if the HV SNe~Ia are systematically fainter than the NV SNe~Ia. Under such circumstances, we would expect that the HV SNe~Ia will be more difficult to find at high-$z$. However, existing evidence suggested that the brightness of HV and NV SNe~Ia are generally comparable \citep[e.g.,][]{2009ApJ...699L.139W,2013Sci...340..170W}. This is also supported by the lack of trend in the relation between \vsiii\ and LC width as noted by many previous studies \citep[e.g.,][]{2009ApJ...699L.139W,2014MNRAS.444.3258M,2021ApJ...923..267D}. These findings make the Malmquist bias less likely to be the underlying source that drives the preference of HV SNe~Ia to be discovered at low-$z$ surveys. However, there is likely a color difference between NV and HV SNe~Ia reported by previous studies. Nonetheless, we would argue the color difference should not be significant enough \citep[$<0.1$\,mag on average;][see also the lower panel of Fig.~\ref{vsi-lc}]{2011ApJ...729...55F} to influence the decision we made to select the Ia-like objects. Moreover, if HV and NV SNe~Ia indeed differ in brightness and color, we should also anticipate a trend from our high-$z$ samples, given their broad redshift coverage (i.e., from $z\sim0.1-0.5$), as opposed to a trend caused solely by samples at $z\lesssim0.1$. A larger high-$z$ sample will be critical for further investigation.  

We next explore the potential impact of different \deltam\ (defined as $B$-band decline 15~days after peak $B$-band brightness) distributions between low-$z$ and high-$z$ surveys. Fig.~\ref{dm15-hist} shows the \deltam\ distributions of various sub-samples in this work. The \deltam\ is converted from the light-curve stretch parameters \citep[e.g.,][]{2007A&A...466...11G,2008ApJ...681..482C} measured by previous studies. It is clear that the \deltam\ distributions of different sub-samples generally cover the \deltam\ range of HV SNe~Ia investigated in this work ($0.71<\deltam<1.55$\,mag, as denoted by the dashed lines in Fig.~\ref{dm15-hist}). However, differences between sub-samples are evident. Notably, the low-$z$ samples (such as PTF and Foundation) exhibit an excess of high-\deltam\ SNe (e.g., $\deltam \gtrsim1.5$\,mag) compared to the high-$z$ samples. To assess if the difference in \deltam\ distributions could impact our results, we construct a new sub-sample by restricting the decline rates of our SNe to $0.7<\deltam<1.6$\,mag (i.e., consistent with the minimum and maximum \deltam\ of our HV SN~Ia sample). Even with this \deltam-restricted sample, the difference in redshift distributions between HV and NV SNe~Ia remains significant. A K-S test gives a $p$-value of $9\times10^{-4}$, rejecting the null hypothesis that the NV and HV SNe~Ia have the same redshift distributions.

\begin{figure}
	\centering
	\begin{tabular}{c}
		\includegraphics*[scale=0.55]{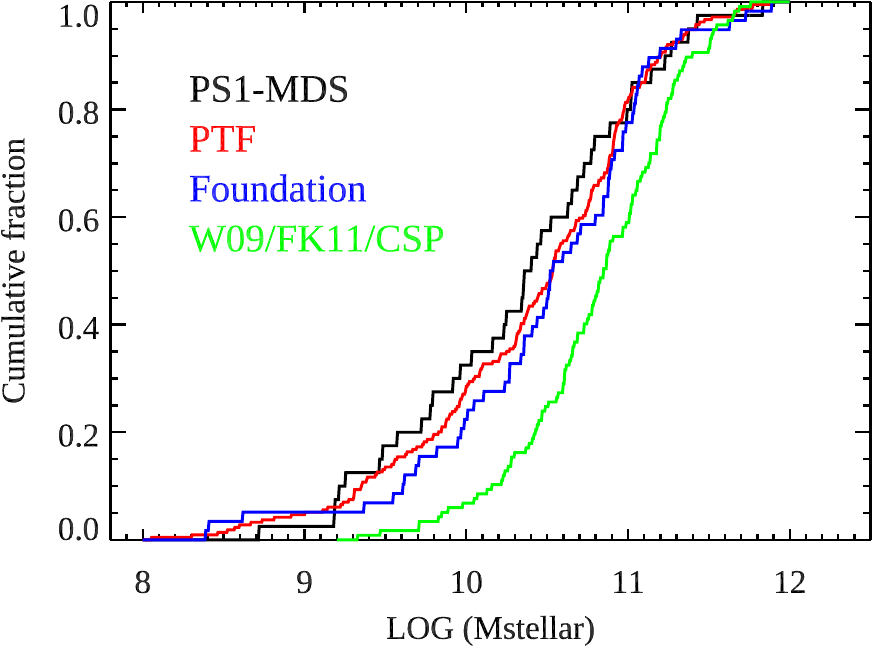}
	\end{tabular}
       \caption{Cumulative fractions of host-galaxy stellar mass (\mstellar) for various samples: PS1-MDS (in black), PTF (in red), Foundation (in blue), and W09/FK11/CSP  (in green) samples. Only those SNe~Ia with \vsiii\ measurements available are included in the figure.
      }
        \label{mass-hist}
\end{figure}

Another potential source of disparity between low-$z$ and high-$z$ surveys could stem from the selection of distinct host-galaxy environments, as HV and NV SNe~Ia tend to reside in galaxies with different \mstellar\ (as discussed in Section~\ref{sec:vsi-host}). Fig.~\ref{mass-hist} compares the distributions of host-galaxy \mstellar\ among various sub-samples. The comparison is made only for those SNe with \vsiii\ measurements available. It is notable that the W09/FK11/CSP samples exhibit a strong bias toward massive galaxies. A K-S test gives a $p$-value of $4\times10^{-4}$, rejecting the null hypothesis that the PS1-MDS and W09/FK11/CSP samples have the same host-galaxy \mstellar\ distributions. This is probably not surprising given that the W09/FK11/CSP samples are mainly galaxy-targeted. In contrast, the \mstellar\ distributions of PTF, Foundation, and PS1-MDS samples are similar to each other (based on the $p$-value of the K-S test). After excluding the W09/FK11/CSP samples from the analysis, the difference in mean redshift between HV and NV subgroups becomes $0.02\pm0.01$ (at 2$\sigma$ significance), and the K-S test gives a $p$-value of 0.02, still rejecting the null hypothesis that the NV and HV SNe~Ia have the same redshift distributions. While it is evident that excluding the SNe from galaxy-targeted surveys greatly reduces the significance of the redshift difference between HV and NV SNe~Ia, it does not eliminate the entire trend. This probably implies the difference in host-galaxy \mstellar\ may not be the only factor contributing to the disparity between low-$z$ and high-$z$ surveys. However, it is still expected to affect the relative fractions between HV and NV SNe~Ia as a function of redshift, as previous observations have shown that the galaxy \mstellar\ functions tend to evolve over time, with galaxies at higher redshift containing on average less \mstellar\ \citep[e.g.,][]{2017A&A...605A..70D}. This trend was also identified with host galaxies of SNe~Ia \citep{2010MNRAS.406..782S}.

While PTF and Foundation samples may have additional complicated selections compared to PS1-MDS, we would argue whether they have a substantial impact on our results, given the similarities of HV and NV SNe~Ia (both photometrically and spectroscopically). In fact, we find that the difference in redshift distributions between HV and NV SNe~Ia persists even when considering solely the PTF and Foundation samples or just the PTF sample. This is against the argument that the discrepancy between low-$z$ and high-$z$ surveys is the main cause of the trend. If this difference is linked to the progenitor properties of SNe~Ia, it may suggest that HV SNe~Ia (and a subset of NV SNe~Ia) tend to have a much longer delay time (i.e., the time between the progenitor star formation and the subsequent SN~Ia explosion) and only becomes significantly prevalent at low-$z$ Universe. To fully understand and disentangle various effects, a larger high-$z$ sample with ejecta velocity measurements will be essential in the future.

\subsection{Implications on SN~Ia cosmology}
\label{sec:cosmology}
\begin{figure}
	\centering
	\begin{tabular}{c}
		\includegraphics*[scale=0.55]{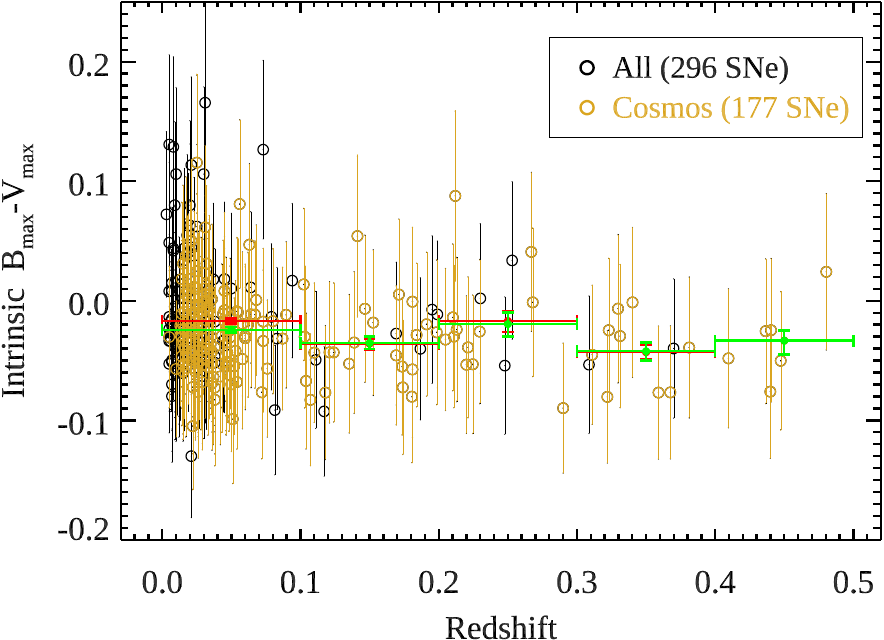}
	\end{tabular}
       \caption{The black open circles show the SN~Ia intrinsic color (as converted from the \Siii\ velocity) as a function of redshift, where the golden open circles represent the SNe only in the cosmological sample. The mean intrinsic color in bins of redshift is shown in red and green solid circles for the original and cosmological samples, respectively.
      }
        \label{color-redshift}
\end{figure}

\begin{figure*}
	\centering
	\begin{tabular}{c}
		\includegraphics*[scale=0.55]{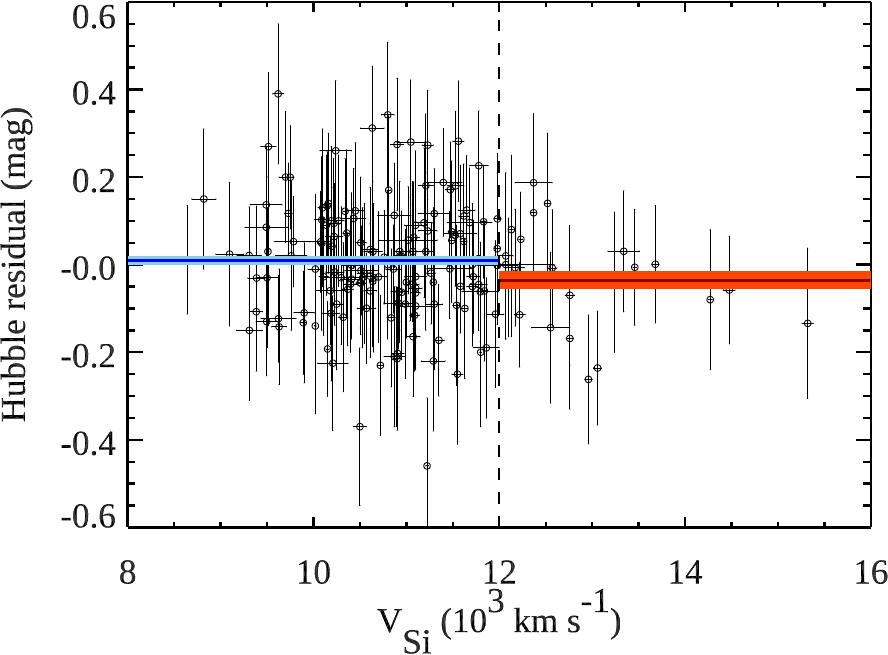}
		\includegraphics*[scale=0.55]{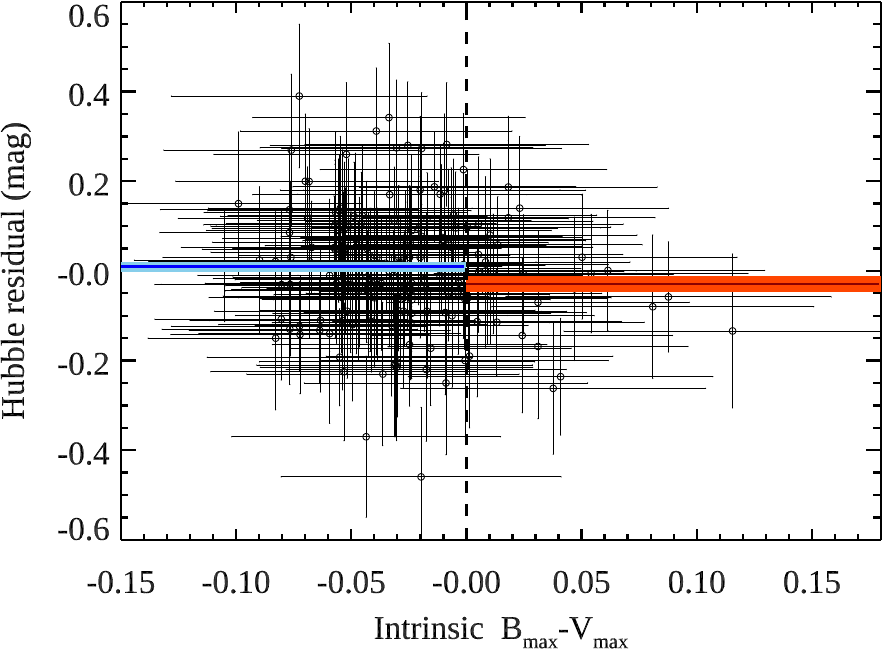}
	\end{tabular}
       \caption{\textit{Left}: The Hubble residuals as a function of \Siii\ velocity (\vsiii). The vertical dashed line represents the criterion used to separate the velocity bins. The solid lines in blue and red represent the weighted mean of the residuals in bins of velocity, and the shaded regions show the 1$\sigma$ uncertainty of the weighted mean. \textit{Right}: The same as the left panel, but with SN~Ia intrinsic color converted from the \vsiii\ using the velocity-color relation \citep{2011ApJ...742...89F}.
      }
        \label{vsi-res}
\end{figure*}

Using a new SN~Ia color model, \citet{2021ApJ...909...26B} suggested that the SN~Ia intrinsic scatter is likely dominated by the $R_V$ variation of the dust around SN. They found the SNe~Ia in massive galaxies have on average lower $R_V$ than those in low-mass galaxies, and this difference can explain the trend between Hubble residuals and host-galaxy \mstellar\ \citep[e.g.,][]{2010MNRAS.406..782S}. They also showed that the SN~Ia color distribution is inconsistent with a symmetric Gaussian distribution, with an excess of red SNe compared with the blue SNe. These red SNe were found to present a larger scatter of distance modulus residuals than the blue SNe.

While \citet{2021ApJ...909...26B} proposed that the host-galaxy correlation with SN~Ia luminosity is due to the variation in dust properties rather than the luminosities of different progenitor systems, it is not clear yet whether this dust variation is intrinsic to the host galaxies (e.g., from the interstellar medium) or the SN progenitor systems. Different SN~Ia explosion scenarios could result in very different dust properties at the SN location. For instance, theoretical studies suggested that the double detonations of sub-Chandrasekhar-mass WDs are likely to have intrinsically different dust properties from that of Chandrasekhar-mass WD explosions \citep[e.g.,][]{2019ApJ...873...84P,2021ApJ...922...68S}. They tend to be redder due to the line blanketing originating from the ashes of the helium shell (although this may also depend on the viewing angle in the model). 

Observationally, \citet{2009ApJ...699L.139W} suggested that the HV SN~Ia prefers a reddening law of lower $R_V$ than that of NV SN~Ia. Assuming most (if not all) of the SNe~Ia in the group-II of bimodal \vsiii\ distribution share the same dust properties as that of HV SNe~Ia, this may qualitatively explain the findings in \citet{2021ApJ...909...26B}, given their preference for lower $R_V$ and massive host galaxies. While the group-II SNe represent only 24\% of the total SNe~Ia in our low-$z$ sample (see Section~\ref{sec:populations}), this fraction will be raised to $\sim$40\% if we only consider the SNe in galaxies of $\log(\mstellar) > 10\,M_{\odot}$. On the other hand, \citet{2018ApJ...859...11S} found that the galaxy dust properties tend to vary with the stellar mass, with massive and quiescent galaxies having lower $R_V$ than that of star-forming galaxies. While HV SN~Ia host galaxies are generally massive, \citet{2015MNRAS.446..354P} found that more than half of the HV SN~Ia host galaxies are not quiescent galaxies. Therefore, the dust from the interstellar medium alone seems challenging to explain the preference of HV SNe~Ia for having a reddening law of lower $R_V$. The nature of the progenitor system could also contribute to the creation of these unique dust properties. If the group-II SNe are associated with the sub-Chandrasekhar-mass explosions, they could be a significant source of distance uncertainties. This is also evident from the previous studies that showed SNe~Ia in low-mass galaxies are better standard candles than those in massive galaxies \citep[e.g.,][]{2019JKAS...52..181K}.

By examining the deviations from a zero-color SN~Ia that is corrected for host-galaxy reddening, \citet{2011ApJ...742...89F} determined the intrinsic $B-V$ color (at peak luminosity) for a sample of SNe~Ia. They found the \vsiii\ is correlated with the SN intrinsic color, with higher-velocity SNe being intrinsically redder than the lower-velocity counterparts (the so-called velocity-color relation). We look into this relation by examining the SN intrinsic color as a function of redshift. We first construct a cosmological sample by including the SNe from our full compilation (i.e., those shown in the right panel of Fig.~\ref{vsi-redshift}) that have Hubble residuals (HRs; defined as the difference between the observed peak SN apparent magnitude and that expected in the assumed cosmological model) measured in \citet{2018ApJ...859..101S} and \citet{2021ApJ...923..267D}. The HRs were measured consistently in both studies. We then compare this sample with the `all' sample, which includes both the cosmological sample and all other SNe that do not meet the cosmological criteria. The SN intrinsic color is converted directly from the \vsiii\ with the velocity-color relation studied in \citet{2011ApJ...742...89F}. As shown in Fig.~\ref{color-redshift}, it is evident that the mean intrinsic color of SNe~Ia is redder for the low-$z$ sample, with the SNe at $z<0.1$ being 0.015\,mag redder than those at $z>0.1$ (at 4.3$\sigma$ significance). However, the trend vanishes ($<2\sigma$) if we only consider the SNe in the cosmological sample. This implies that the difference in intrinsic color (as converted from velocity) between low-$z$ and high-$z$ samples can be effectively removed by the conventional cosmological cuts.

We next investigate the potential distance bias in cosmology by examining the relation between HR and \vsiii.  \citet{2020MNRAS.493.5713S} found a possible trend (in 2.7\,$\sigma$) between \vsiii\ and HR, in the sense that higher-velocity SNe~Ia tend to show negative average HRs, while lower-velocity SNe~Ia tend to show positive average HRs. However, this trend becomes insignificant if using the SNe mainly discovered by un-targeted surveys \citep{2021ApJ...923..267D}. The HR as a function of \vsiii\ and SN intrinsic color with our sample are shown in Fig.~\ref{vsi-res}. This is the same cosmological sample as that shown in Fig.~\ref{color-redshift}. We find that the HV and redder (i.e., with positive intrinsic color) SNe tend to show negative average HRs, while their counterparts tend to show positive average HRs. Although our results are generally consistent with the previous studies, the trends are not statistically significant, with differences of $0.044\pm0.023$\,mag (1.9$\sigma$) and $0.038\pm0.021$\,mag (1.8$\sigma$) between the sub-groups in HR-\vsiii\ and HR-color relations, respectively. Nevertheless, we note that the current number of HV SNe~Ia in our cosmological sample is considerably smaller than that of NV SNe~Ia. 

\section{Conclusions}
\label{sec:conclusions}
In this paper, we measure the \Siii\ velocity (\vsiii) from SNe~Ia discovered by the PS1-MDS and investigate the relations between \vsiii\ and other parameters, including LC width, color, and host-galaxy properties. Our main findings are:

\begin{enumerate}

\item[$\bullet$] We do not see significant trends between \vsiii\ and either LC width $x_1$ or color $c$. For the relation with host-galaxy \mstellar, it is difficult to claim the trend with the PS1-MDS sample alone, given its small sample size. However, we note that the trend stays significant after combining our PS1-MDS SNe with those from low-$z$ studies, in the sense that HV SNe~Ia are more likely to be found in massive galaxies.

\item[$\bullet$] No significant trends are identified between \vsiii\ and redshift using either the PS1-only sample or a large spectroscopic compilation that includes additional low-$z$ and high-$z$ datasets. However, with the large sample, we find a significant probability that HV and NV SNe~Ia may have distinct redshift distributions, with HV SNe~Ia being more prevalent in the local Universe. Different selections between low-$z$ and high-$z$ surveys could play a role. Nevertheless, we do not rule out the possibility that the progenitors of HV SNe~Ia (and perhaps a subset of NV SNe~Ia) may favor explosion scenarios of long delay times.

\item[$\bullet$] We confirm the previous findings that the \vsiii\ distribution of low-$z$ SNe Ia shows a significant bimodal distribution. The two groups substantially overlap in the velocity space, with a fraction of 76\% and 24\% for SNe in group-I and group-II, respectively. However, this bimodality becomes insignificant for the high-$z$ sample studied in this work.

\item[$\bullet$] Our results are also in line with recent cosmological studies, which have shown that SNe~Ia in massive galaxies typically exhibit lower $R_V$ than those in low-mass galaxies. This could possibly be explained by assuming that a large fraction of SNe~Ia in massive galaxies share similar dust properties with HV SNe~Ia. 

\item[$\bullet$] By converting the \vsiii\ to the SN intrinsic color, we show that the mean intrinsic color of SNe~Ia appears to be redder in the local Universe. However, this trend vanishes when considering only the SNe in the cosmological sample. Furthermore, we confirm that the HV SNe~Ia tend to have negative average HRs than their counterparts, but the trend is not statistically significant with our sample.
\end{enumerate}

Future studies with precise spectroscopic measurements spanning a wide range of redshift would be crucial to differentiate between SN~Ia explosion scenarios and investigate their potential evolution or change in demographics. Such investigations will enormously improve our understanding of the nature of SNe~Ia and refine their utility in cosmological studies.

\section*{acknowledgments}
This work is supported by the National Science and Technology Council (NSTC grant 109-2112-M-008-031-MY3).

\section*{Data Availability}
The SN~Ia spectra from PS1-MDS will be made available in the WISeREP archive \citep{2012PASP..124..668Y}. Other data underlying this article will be shared on reasonable request to the corresponding author.



\bibliographystyle{mnras}
\bibliography{ps1_spec} 






\appendix
\section{The \Siii\ feature of the PS1-MDS \vsiii\ sample in this work}

\begin{figure*}
	\centering
	\begin{tabular}{c}
		\includegraphics*[scale=0.85]{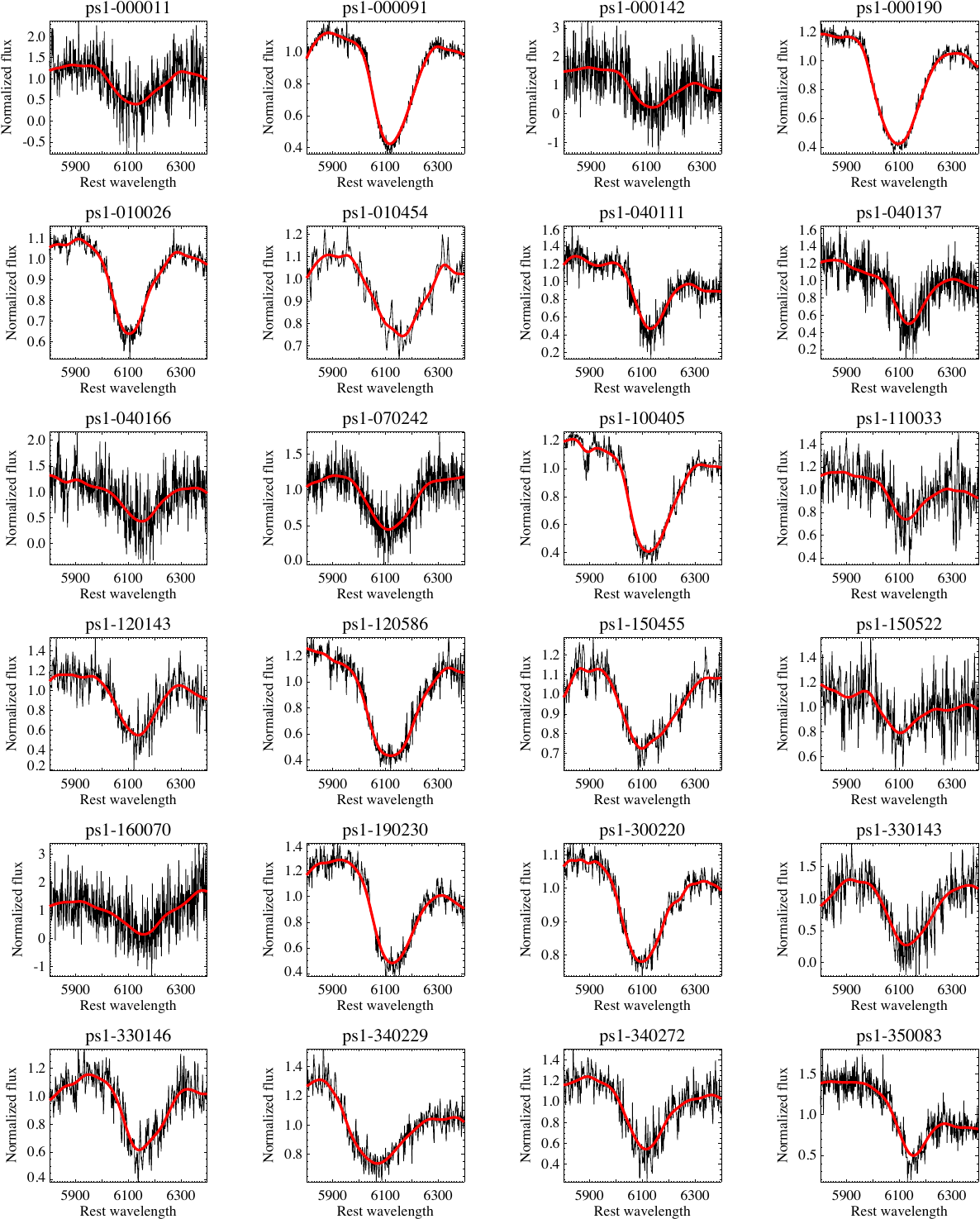}
	\end{tabular}
       \caption{The \Siii\ absorption feature measured in this work. The observed and smoothed spectra are shown in black and red, respectively. 
      }
        \label{vsi_mosaic1}
\end{figure*}

\begin{figure*}
	\centering
	\begin{tabular}{c}
		\includegraphics*[scale=0.85]{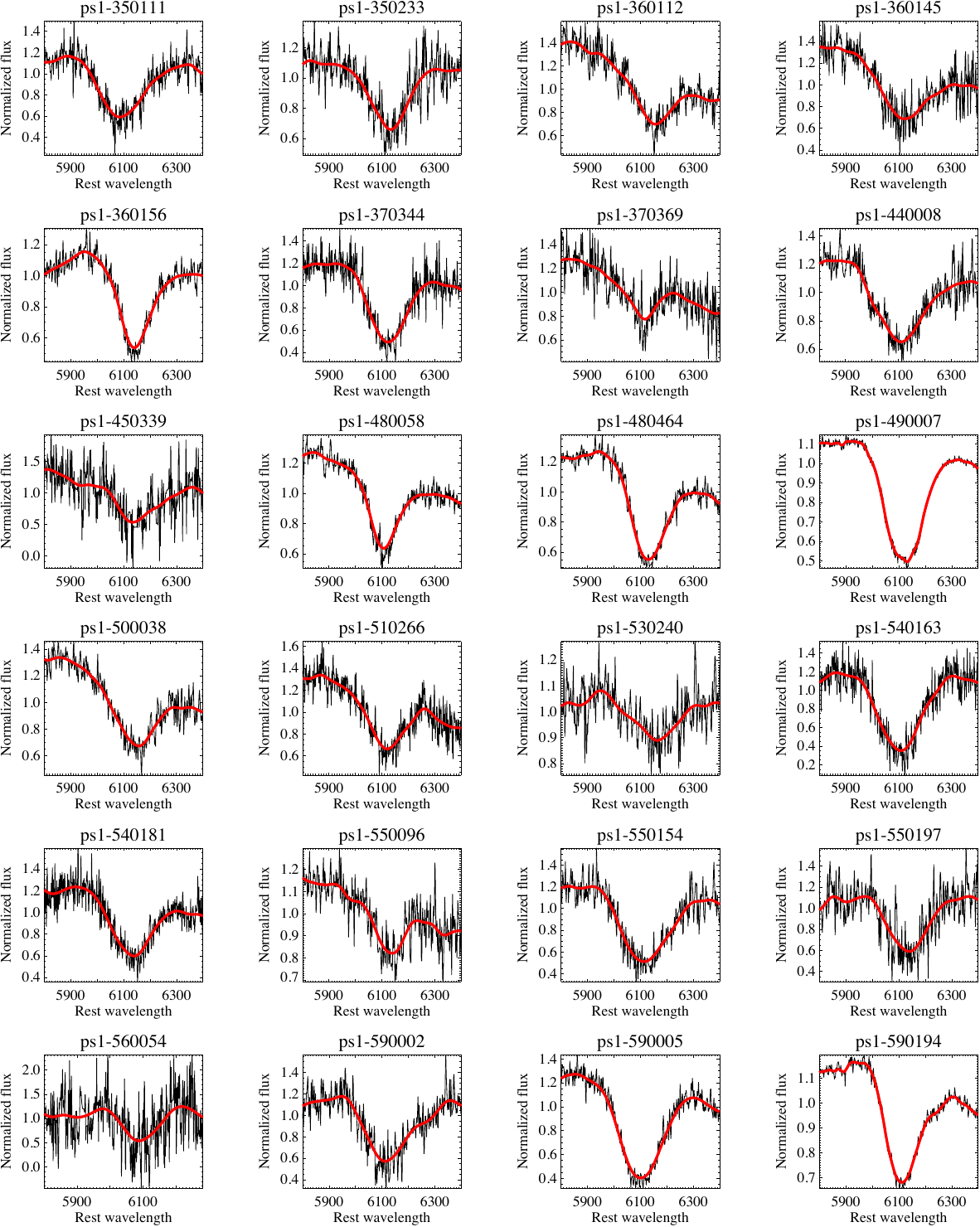}
	\end{tabular}
       \caption{The \Siii\ absorption feature measured in this work. The observed and smoothed spectra are shown in black and red, respectively (continued).
      }
        \label{vsi_mosaic2}
\end{figure*}

\newpage
\section{The summary of \vsiii\ measurements in this work}

\begin{table*}
\centering
\caption{The \Siii\ velocity (\vsiii) measurements of high-$z$ samples (including PS1-MDS, SDSS, and SNLS) used in this work.}
\label{hiz_vsi}
\scriptsize
\begin{tabular}{lcrrr}
\hline\hline
SN Name & Redshift & Phase & \vsiii\ & $\sigma_{\vsiii}$ \\
        &  ($z$)   & (days) & (\kms) & (\kms) \\     
\hline
PS1-000011 & 0.381 & 2.7    & $-10633$ & 128 \\
PS1-000091 & 0.153 & 2.7    & $-11265$ & 47 \\
PS1-000142 & 0.437 & 1.6    & $-11048$ & 151 \\
PS1-000190 & 0.102 & $-2.6$ & $-12232$ & 19 \\
PS1-010026 & 0.032 & $-3.4$ & $-11831$ & 28 \\
PS1-010454 & 0.081 & 2.0    & $-9048$  & 55 \\
PS1-040111 & 0.370 & 4.3    & $-10609$ & 40 \\
PS1-040137 & 0.447 & 2.8    & $-10298$ & 214 \\
PS1-040166 & 0.440 & 3.5    & $-9515$  & 82 \\
PS1-070242 & 0.064 & $-4.5$ & $-11473$ & 51 \\
PS1-100405 & 0.103 & 4.8    & $-10907$ & 69 \\
PS1-110033 & 0.205 & $-3.5$ & $-10836$ & 45 \\
PS1-120143 & 0.174 & 3.2    & $-10152$ & 19 \\
PS1-120586 & 0.230 & 0.9    & $-11039$ & 58 \\
PS1-150455 & 0.064 & $-0.4$ & $-12157$ & 44 \\
PS1-150522 & 0.230 & 2.3    & $-11878$ & 68 \\
PS1-160070 & 0.359 & 3.2    & $-9491$  & 171 \\
PS1-190230 & 0.139 & 2.3    & $-10760$ & 40 \\
PS1-300220 & 0.094 & $-3.2$ & $-12329$ & 132 \\
PS1-330143 & 0.187 & 0.8    & $-10600$ & 265 \\
PS1-330146 & 0.135 & 1.4    & $-10223$ & 92 \\
PS1-340229 & 0.141 & 4.7    & $-13459$ & 19 \\
PS1-340272 & 0.195 & $-0.9$ & $-11595$ & 194 \\
PS1-350083 & 0.368 & 4.7    & $-9496$  & 113 \\
PS1-350111 & 0.253 & 2.4    & $-12841$ & 215 \\
PS1-350233 & 0.169 & $-3.6$ & $-10435$ & 126 \\
PS1-360112 & 0.181 & $-0.5$ & $-9388$  & 96 \\
PS1-360145 & 0.199 & 4.9    & $-11020$ & 314 \\
PS1-360156 & 0.046 & $-4.0$ & $-10147$ & 14 \\
PS1-370344 & 0.331 & 2.9    & $-10923$ & 89 \\
PS1-370369 & 0.191 & 4.1    & $-11223$ & 17 \\
PS1-440008 & 0.147 & $-1.6$ & $-11617$ & 23 \\
PS1-450339 & 0.410 & $-0.5$ & $-10359$ & 42 \\
PS1-480058 & 0.034 & 4.9    & $-11739$ & 28 \\
PS1-480464 & 0.221 & $-1.1$ & $-10638$ & 15 \\
PS1-490007 & 0.055 & 0.8    & $-11525$ & 21 \\
PS1-500038 & 0.107 & 4.7    & $-9307$  & 130 \\
PS1-510266 & 0.440 & $-3.4$ & $-11072$ & 76 \\
PS1-530240 & 0.117 & $-0.2$ & $-9015$  & 51 \\
PS1-540163 & 0.330 & $-4.6$ & $-11619$ & 56 \\
PS1-540181 & 0.309 & $-4.7$ & $-10199$ & 93 \\
PS1-550096 & 0.225 & $-4.2$ & $-10209$ & 164 \\
PS1-550154 & 0.210 & $-1.3$ & $-11397$ & 199 \\
PS1-550197 & 0.248 & 3.6    & $-10175$ & 130 \\
PS1-560054 & 0.480 & $-0.2$ & $-12553$ & 206 \\
PS1-590002 & 0.199 & $-1.4$ & $-11473$ & 68 \\
PS1-590005 & 0.171 & 0.2    & $-11978$ & 28 \\
PS1-590194 & 0.090 & $-1.8$ & $-11463$ & 15 \\
SDSS2372   & 0.181 & 2.1    & $-11795$ & 197 \\
SDSS2561   & 0.118 & 0.0    & $-9492$  & 229 \\
SDSS2789   & 0.290 & 3.9    & $-9097$  & 151 \\
SDSS2916   & 0.124 & 1.2    & $-10509$ & 53 \\
SDSS3080   & 0.174 & $-4.5$ & $-9623$  & 189 \\
SDSS3592   & 0.087 & 0.2    & $-10861$ & 84 \\
SDSS5533   & 0.220 & 2.3    & $-10199$ & 34 \\
SDSS5549   & 0.121 & 0.3    & $-10512$ & 174 \\
SDSS6057   & 0.067 & 3.4    & $-11469$ & 249 \\
SDSS6315   & 0.267 & 2.7    & $-13058$ & 44 \\
SDSS6699   & 0.311 & 3.7    & $-10450$ & 86 \\
SDSS6933   & 0.213 & 0.4    & $-11089$ & 52 \\
SDSS6936   & 0.181 & $-0.3$ & $-10077$ & 121 \\
SDSS7147   & 0.110 & 0.4    & $-10503$ & 37 \\
SDSS7475   & 0.322 & 3.1    & $-9384$  & 68 \\
SDSS7847   & 0.212 & $-1.4$ & $-14476$ & 63 \\
SDSS10434  & 0.104 & 2.5    & $-9785$  & 202 \\
04D1dc     & 0.211 & $-1.7$ & $-10911$ & 155 \\
04D3nh     & 0.340 & 4.5    & $-11778$ & 221 \\
05D2ab     & 0.323 & $-1.3$ & $-11072$ & 65 \\
05D2ah     & 0.184 & $-2.1$ & $-10951$ & 98 \\
05D3ne     & 0.169 & $-4.0$ & $-10987$ & 110 \\
06D3fp     & 0.268 & $-0.1$ & $-11779$ & 90 \\
\hline
\end{tabular}
\end{table*}






\bsp	
\label{lastpage}
\end{document}